documentclass [prb,preprint,showpacs]{revtex4}
\documentclass [prb,twocolumn]{revtex4}
\usepackage{amssymb}
\usepackage{amsmath}
\usepackage{epsfig}

\begin{document}
\title{Critical behavior of the three-dimensional compressible Ising antiferromagnet at constant volume: a Monte Carlo study}
\author{Luigi Cannavacciuolo and D. P. Landau}
\affiliation{Center for Simulational Physics, The University of Georgia,
Athens, GA 30602 USA}
\date{\today}
\begin{abstract}
Extensive Monte Carlo simulations in the semi-grand-canonical ensemble
are used to study the critical behavior of a three-dimensional compressible Ising model with antiferromagnetic interactions under constant volume conditions. Elastic forces between spins are introduced by the Stillinger-Weber potential and energy parameters are chosen in such a way that antiparallel spin ordering is favored, analogous to the antiferromagnetic coupling in the rigid Ising Hamiltonian. All the quantities analyzed strongly indicate that the system remains in the universality class of the standard (rigid) three-dimensional Ising model, in contrast with theoretical predictions.
\end{abstract}
\maketitle
\section{Introduction}The Ising model, first proposed by Lenz in 1925 as a microscopical model for ferromagnetism, has a (constant) interaction $J$ included only between the nearest neighbors spins on a (rigid) lattice. Despite its (formal) simplicity an analytical solution has been found only for one-, and two-dimensional models in zero magnetic field. \cite{onsager} For the three-dimensional Ising model only approximate solutions are available (e.g. series expansions, \cite{serie_exp} renormalization group calculations \cite{rg}, $\epsilon$-expansions, \cite{eps_ising} Monte Carlo renormalization group calculations \cite{mcrg}, and Monte Carlo (MC) simulations), \cite{ising_mc} although quite precise results have been found. 

It was clear, a few years after its introduction, that the Ising model could be employed to describe other phase transitions. For instance, phase separation in binary alloys can be studied by an Ising model in which \lq\lq up\rq\rq and \lq\lq down\rq\rq  spins are replaced with sites occupied by a \lq\lq A\rq\rq or \lq\lq B\rq\rq type atom, respectively.  
The chemical potential plays the role of the magnetic field, the density that of the magnetization, etc. and the appropriate statistical ensemble is the semi-grand-canonical, instead of the canonical.
With the above mentioned analogy in mind, in the present work we will use the languages of magnetism and of alloys interchangeably. 

In a real magnetic crystal, however, atoms interact with each other through a combination of elastic and magnetic forces. The next step in realism is, therefore, the explicit introduction of elastic degrees of freedom in the traditionally rigid system. The resulting model is termed {\em compressible Ising} (CIM). Several empirical potentials have been proposed and employed to mimic the elastic force, from the simple Lennard-Jones to the more specific Tersoff, \cite{tersoff} Keating, \cite{keating} Stillinger-Weber, \cite{sw} and others (these last three have been introduced, mainly to reproduce the interaction between silicon and germanium atoms in the study of Si$_{1-x}$Ge$_x$, viewed as Ising binary alloy models).

The issue of how the presence of elastic interactions affects the critical behavior of the Ising model has been intensively studied. In 1954, using thermodynamic considerations, Rice \cite{rice} predicted that if an Ising system with divergent specific heat is put on a deformable lattice at constant pressure, it undergoes a first order transition. A number of calculations appeared later \cite {domb,mattis,garland}, all of which lead to the conclusion that a first order transition was expected to occur. All these studies assume that a \lq\lq pure\rq\rq Ising singularity happens at constant volume. In 1968 Fisher \cite{fisher_hidden} changed the situation drastically with his \lq\lq hidden variable\rq\rq theory, assuming that the \lq\lq pure\rq\rq phase transition occurs at fixed {\em intensive} variable, i.e. at fixed pressure. As a result of this hypothesis a second order transition was found with Fisher renormalized exponents, if the critical exponent of the specific heat $\alpha$ is positive. A significant highlight in this controversy was the work of Larkin and Pikin \cite{lp} who, for the first time, considered a Hamiltonian with fluctuations of both the order parameter and the elastic modes and pointed out the special role of the macroscopic mode {\bf k=0} (in Fourier space). As a result a first order transition is found at constant pressure and quadratic coupling of the order parameter and the strain tensor, also if $\alpha=0$. It was later recognized that this result is only valid at low pressure, whereas at high pressure the critical behavior seems to be much more complex. \cite{imry,belim2001} A different approach was used by Baker and Essam \cite{be} who mapped an Ising model on a compressible lattice, including the coupling between magnetic and elastic degrees of freedom,  onto a standard Ising (on a rigid lattice), but with parameters which depend on the elastic degrees of freedom. At constant volume the model exhibited identical critical behavior as the underlying rigid system, and not a first order transition as predicted. \cite{lp} At constant pressure a second order transition was found with Fisher renormalized exponents. This model, as well as others of the same type, however, considers a negligible shear modulus. This somewhat unphysical assumption was soon recognized to be the reason of the disappearance of the first order transition. 
Further work \cite {altri1,altri2,altri3,altri4} with more elaborate schemes did not change the result: a first order transition was always predicted. Finally, we cite the long and complicated work by Bergman and Halperin, \cite{bergman} in which a three-dimensional anisotropy in the elastic forces was introduced. In the isotropic case the same already established result was obtained, i.e. a first order transition at constant pressure if $\alpha > 0$, and a second order transition with renormalized exponents at constant volume. However, this last condition is realized only if the atoms at the surface are fixed, otherwise the system has still enough degrees of freedom to develop a macroscopic instability. In the anisotropic case, however, a so-called microscopic instability was found at constant volume, and this is interpreted as a first order transition. In his Habilitation thesis D\"unweg \cite{burk} completely revised the topic. Starting from the Landau-Ginzburg-Wilson Hamiltonian he was able to show that the CIM under constant pressure exhibits mean-field critical behavior, \cite{cowley} in agreement with simulations. \cite{dl_mf,mohamed} At constant volume he predicted two first order lines ending in critical points, which are likely to belong to the mean-field universality class. If the magnetic interactions are {\em antiferromagnetic} (AFM) instead of ferromagnetic, \cite{ipotesi_burk} a quadratic coupling between the order parameter and the elastic deformation should be expected. In this case the predictions are: (i) a second order phase transition with Fisher-renormalized exponents in the case of constant volume \cite{nota_afcv} (this is in agreement with the $\epsilon$-expansion work of Bergman and Halperin \cite{bergman}); (ii) a first order phase transition in the case of antiferromagnetic interactions and constant (zero) pressure. \cite{nota_afcp} As pressure increases, the first order line in pressure-temperature space should split into two first order lines at a triple point. The theoretical prediction in the case of FM interactions at constant volume has been checked by Tavazza {\em et al.} \cite{francesca} by MC simulations. In disagreement with theory, they found a closed first order line which separates ordered and disordered phases.

This result raises the intriguing question of whether or not theory will be correct for the case of AFM interactions and constant volume. 
This will be the topic of the study that we report here.

\section{Model and simulation techniques}
We use the same model as considered previous in Ref. \onlinecite {mohamed,francesca}. We consider a binary alloy of A and B atoms on the nodes of a distortable diamond lattice, free to move on the condition that the diamond four-fold coordination is preserved, and that the atomic species on each node may change. The coordination requirement speeds up considerably the simulations as the list of the nearest neighbors of a given atom, which enter the Hamiltonian (see below), is known from the very beginning and does not change during the simulations. The diamond “lattice” is decomposed into eight interpenetrating simple cubic sublattices of linear size $L$, so there  are $N=8L^3$ atoms (sites). 
Each atom in the system has four degrees of freedom: the three spatial coordinate {\bf r} and its species S, which is defined to be $S=1$  if the atom is an A, or $S=-1$ if it is a B type. The total number of atoms $N$ is kept constant during the simulation, while the relative concentrations of A and B particles can vary and are controlled by the chemical potential. The corresponding appropriate statistical ensemble is termed semi-grand canonical. The Hamiltonian is given by
\begin{equation}
H = -\frac{1}{2}(\mu_{\rm A}-\mu_{\rm {B} })\sum_i S_i + H_{SW},
\end{equation}
where $\mu_{\rm A}$, $\mu_{\rm {B} }$ are the chemical potentials 
of A and B, respectively, and $H_{SW}$ is the Stillinger-Weber (SW) potential given by
\begin{equation} 
H_{SW} = H_{2bd} + H_{3bd}.
\end{equation}
The SW potential contains a two body interaction $H_{2bd}$ involving nearest neighbors and a three body $H_{3bd}$ interaction that includes next-nearest neighbors as well. This last term is essential to stabilizing a diamond-like structure. The two body interaction is given by
\begin{equation}
H_{2bd} = \sum_{\langle i,j \rangle} \epsilon(S_i,S_j) F_2[r_{ij}/R_0(S_i,S_j)],
\end{equation}
where $R_0$ is the ideal distance of the atoms, $r_{ij}$ is the distance between sites $i$ and $j$, $F_2$ is a short range function of the rescaled bond length (see Ref. \onlinecite{mohamed} for details), and $\epsilon$ is the covalent binding energy. For a binary alloy of silicon and germanium (Si=A, Ge=B) it has been estimated $\epsilon(1,1)$=2.17 eV, $\epsilon$(-1,-1) = 1.93 eV, and $\epsilon(1,-1)$ = 2.0427 eV, \cite{mohamed} and these values were used for of an elastic ferromagnet simulations. \cite{mohamed,francesca} In the present work we increased the value for the A-B binding energy by 0.3 eV to favor alternate ordering of A-B particles analogous to the antiferromagnetic ordering in magnets, specifically $\epsilon(1,-1)= 2.3427$. The behavior for this value is expected to be typical of that in the AF regime, but a much larger value could conceivably produce unanticipated effects. A systematic study of the dependence upon the value of  $\epsilon(1,-1)$ is beyond the scope of this paper.  
The three body part of the Hamiltonian is given by
\begin{eqnarray}
&&H_{3bd} = \sum_{\langle i,j,k \rangle}\{\epsilon(S_i,S_j)^{1/2} 
\epsilon(S_j,S_k)^{1/2}\mathcal{L}(S_i,S_j,S_k) \times \nonumber \\
&& F_3[r_{ij}/R_0(S_i,S_j), r_{jk}/R_0(S_j,S_k)] (\cos \theta_{ijk}+\frac{1}{3})^2\},
\end{eqnarray}
where $F_3$ is a function of the same kind of $F_2$, $\mathcal{L}$ is a simple function of the atomic species (see Ref. \onlinecite{mohamed} for details), and $\cos \theta_{ijk} = {\bf r}_{ij} \cdot {\bf r}_{jk}/|{\bf r}_{ij} \cdot {\bf r}_{jk}|$. The sum is performed over all triplets (i,j,k) with the vertex at site $j$, $i$ and $k$ are nearest neighbors of $j$. Note that $H_{3bd}$ contains an angular term which is a sort of angular stiffness which is essential to stabilizing the diamond lattice (in fact, assigning the bonds length alone is not sufficient, because the lattice has 3$N$ translational degrees of freedom, and the bond length only imposes 2$N$ constraints). A single MC step is performed in the following manner: an atom of species $S_i$ at position ${\bf r}_i$ is randomly chosen and a transition to the state ${\bf r}'_i$, $S'_i$ is attempted. The change in energy is then calculated and the move is accepted or rejected according to the standard Metropolis criterion. Note that not necessarily both ${\bf r}'_i$ and $S'_i$ have to be different from the initial values ${\bf r}_i$, $S_i$. We simulated systems of sizes $L$ = 4 ($N$=512), 6 ($N$=1728), 8 ($N$=4096), 10 ($N$=8000), 12 ($N$=13824), 14 ($N$=21952), and 18 ($N$=46656). 
Periodic boundary conditions were used. A number ranging from $5\times 10^4$ MCS for the smallest systems to $5\times 10^5$ MCS were discarded to thermalize the system. The typical number of MC steps for sampling ranges from $2\times 10^6$ to $6\times 10^6$. For each system size we performed 10-50 independent runs, using different random number sequences to achieve a satisfactory statistical error on the averages of the sampled quantities. During a simulation the volume is kept constant at a value corresponding to the lattice constant $a_0$ of the ferromagnet, for consistency to the simulations of Refs. \onlinecite{mohamed, francesca}. 
Simulations were performed at fixed chemical potential $\mu_{\rm B} = \mu_0$ varying the temperature $T$, as well as at fixed $T=T_0$ varying $\mu_{\rm B}$ (see Sec. \ref{mc_result}). 
In this study we deal with antiferromagnetic ordering and consider atoms sitting on an \lq\lq even\rq\rq site to belong to a sublattice SL$_1$ and those sitting on an \lq\lq odd\rq\rq site to belong to a different sublattice SL$_2$. 
During the simulations we sampled the SW energy, and the fraction of  A and B particles in SL$_1$ and SL$_2$. Using these quantities as input, we calculated all the thermodynamic quantities needed by employing the histogram reweighting method. \cite{isto}

The order parameter $M$ is  defined by the absolute value of the staggered magnetization $m^+$, i.e.
\begin{equation}
 m^+ =  \frac{1}{2} \left (  \sum_{SL_1}S_i - \sum_{SL_2}S_i \right ),
\label{m+}
\end{equation}
and
\begin{equation}
M = \langle |m^+| \rangle
\end{equation}
where the first and second sums in Eq. (\ref{m+}) are performed over spins belonging to the SL$_1$ and SL$_2$ sublattices, respectively. The finite lattice staggered susceptibility of the order parameter is
\begin{equation}
\chi^+_{fl} = N \left (\langle m^+{}^2 \rangle - \langle |m^+| \rangle^2 \right )/k_B T,  
\end{equation}
where $k_B$ is the Boltzmann constant and $T$ the absolute temperature; 
the staggered susceptibility is
\begin{equation}
\chi^+ = N \left (\langle (m^{+})^{2} \rangle - \langle m^+ \rangle^2 \right )/k_B T.  
\end{equation}
We also determined the average concentration of the B species
\begin{equation}
c_B = \frac{1}{2N} \langle n_B \rangle, 
\end{equation}
where $n_B$ is the number of B particles in the system; 
the reduced fourth order cumulant of the order parameter
\begin{equation}
U_4 = 1 - \frac{\langle m^+{}^4  \rangle} {3 \langle m^+{}^2 \rangle^2};
\end{equation}
the specific heat
\begin{equation}
C_v = N (k_B T)^2 \left ( \langle E^2 \rangle - \langle E \rangle^2 \right ),
\end{equation}
where $E$ is the total energy. In our analysis we considered, in addition, the logarithmic derivatives of $m^+$ and of $U_4$ with respect to $T$. To calculate them we used the relation
\begin{equation}
\frac{\partial}{\partial T} \ln \langle |m^+|^n \rangle =  
\frac{\langle |m^+|^n E \rangle}{\langle |m^+|^n \rangle} - \langle E \rangle, \;\mbox{$n = 1,2,\dots$}.
\label{deriv_log}
\end{equation}
Moreover, the derivatives of a thermodynamic quantity $X$ with respect to $T$ were calculated using its cross-correlation with the energy, i.e.
\begin{equation}
\frac{ \partial |X| }{\partial T} = \langle |X| E \rangle -\langle |X| \rangle \langle E \rangle.
\label{deriv_td}  
\end{equation}    

\section{Monte Carlo results}
\label{mc_result}
\subsection{Phase diagram}
\begin{figure}[h]
\begin{center} 
\mbox{
\epsfxsize=6cm 
\epsfysize=5cm 
\epsffile{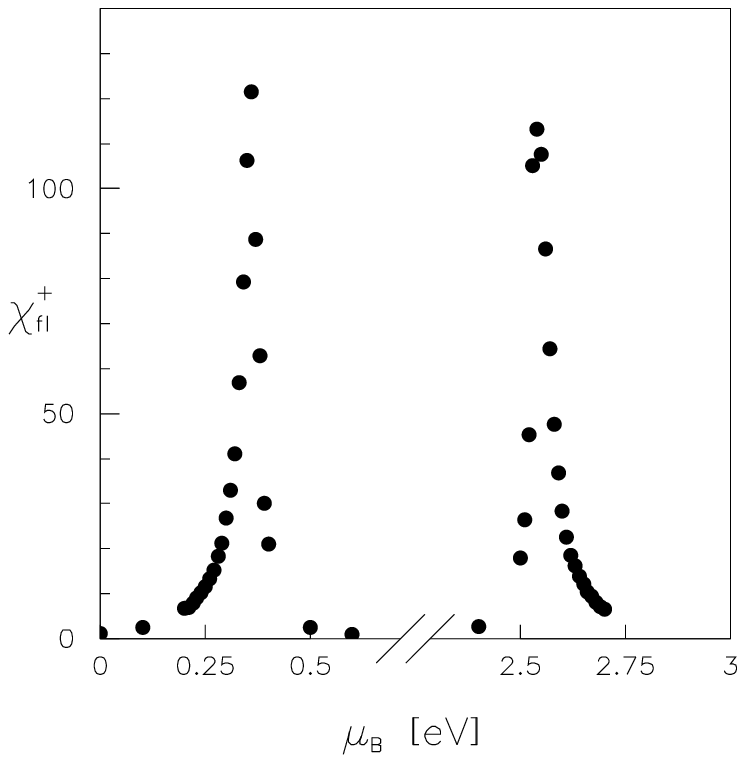}
}
\mbox{
\epsfxsize=6cm 
\epsfysize=5cm 
\epsffile{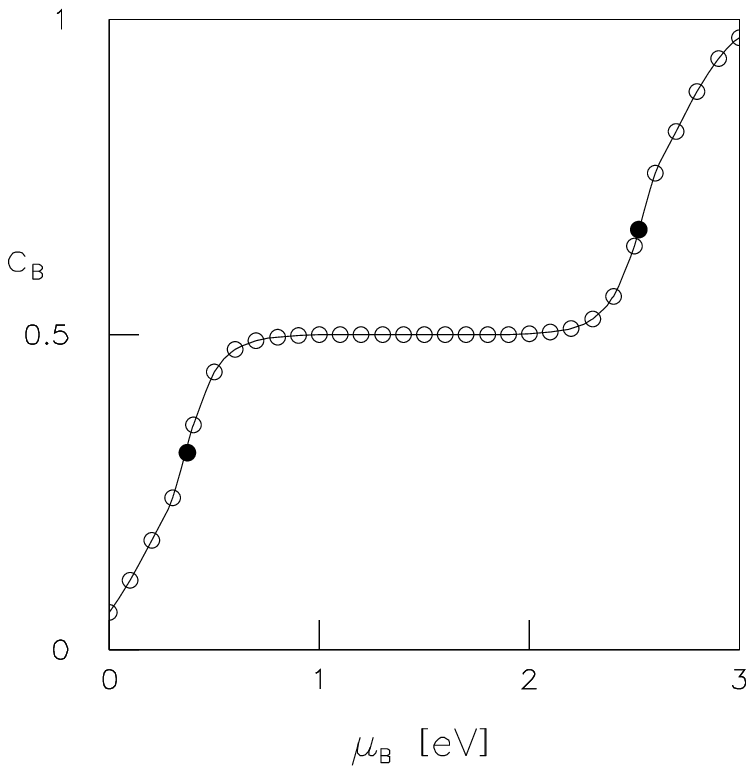}
}
\end{center}
\caption{Typical data used to determine the critical points. 
(a) Plot of $\chi^+_{fl}$ vs $\mu_{\rm B}$. (b) Concentration of the B species vs $\mu_B$. The bold circles show the location of the transitions. $L=4$ and the temperature is $T=0.1$ eV in both plots. The error bars are smaller than the size of symbols.}
\label{raw_data}
\end{figure}

\begin{figure}[h]
\begin{center} 
\mbox{
\epsfxsize=6cm 
\epsfysize=5cm 
\epsffile{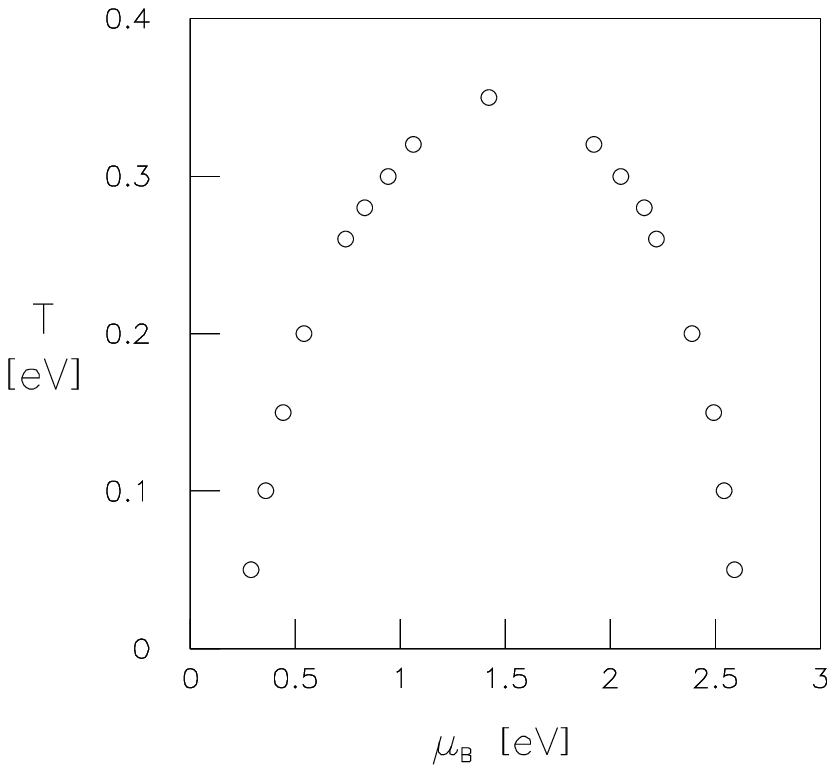}
}
\mbox{
\epsfxsize=6cm 
\epsfysize=5cm 
\epsffile{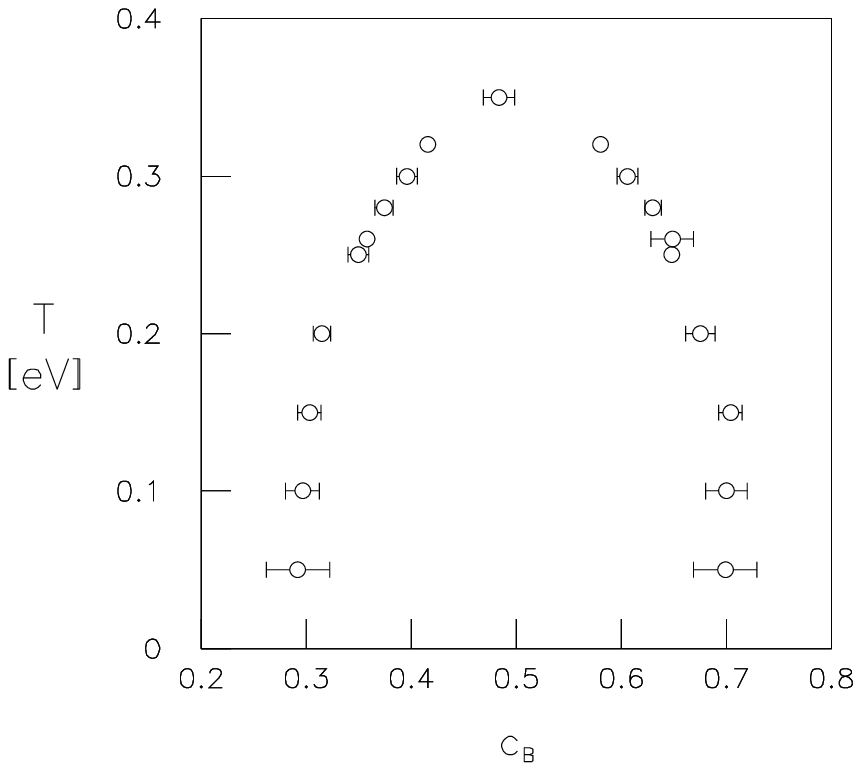}
}
\end{center}
\caption{Phase diagram of the AF compressible Ising model at constant volume in the $(\mu_{\rm B},T)$ plane (a), and  in the $(c_{\rm B},T)$ plane (b). The second order line separates the ordered-disordered phases. Estimates are for $L=4$. Note the slight asymmetry of the curve which reflects the asymmetry of our model. The error bars if not visible are smaller than the size of symbols.}
\label{ph_dia}
\end{figure}

We started the present study with a rough determination of the phase diagram in the $(\mu_{\rm B},T)$ plane. We performed simulations at different values of $T$ ranging from 0.05 eV to 0.35 eV. \cite{nota_t} At fixed $T$ we swept the chemical potential $\mu_{\rm B}$ from 0 to 4 eV at intervals of 0.1 eV, while $\mu_{\rm A}$ was kept constant at 1 eV. For each value of $T$ we determined the value of $\mu_{\rm B}$ at which the maximum value of $\chi$ occurs (see Fig. \ref{raw_data}(a)).
The behavior of concentration $c_B$ as the chemical potential is swept is shown in Fig. \ref{raw_data}(b). The solid dots indicate the locations of the peaks in the finite lattice ordering susceptibility. The resulting phase boundary is plotted in Fig. \ref{ph_dia}(a), whereas Fig. \ref{ph_dia}(b) shows the phase diagram in the $(c_{\rm B},T)$ plane. 
Note that this procedure gives only an approximate phase diagram, which does not take into account extrapolation to the thermodynamic limit. It is, however, very useful as it provides us with information on where to focus in the $(\mu_{\rm B},T)$ plane with higher resolution simulations and with finite size scaling analysis. A transition point from the disordered to the ordered phase is estimated at $\mu_{\rm B} \simeq 1.42$ eV and $T \simeq 0.34$ eV. We decided, therefore, to keep $\mu_{\rm B}$ fixed at the above value and to run simulations in a neighbor of $T \simeq 0.34$ eV. This corresponds to moving along the y-axis, i.e. perpendicularly to the phase diagram of Fig. \ref{ph_dia}(a). We also performed simulations moving parallel to the phase diagram at $T=0.1$ eV and varying $\mu_{\rm B}$. The results, not shown in this paper, are consistent with those obtained by moving in the orthogonal direction. Note the slight asymmetry of the curves in Fig. \ref{ph_dia} which is a consequence of the three body interactions. The phase boundary shows no hysteresis, and a finite size scaling analysis (descibed in the next section) indicates that the transition is 2nd order.

\subsection{Critical behavior}
\begin{figure}[h]
\begin{center} 
\mbox{
\epsfxsize=6cm 
\epsfysize=5cm 
\epsffile{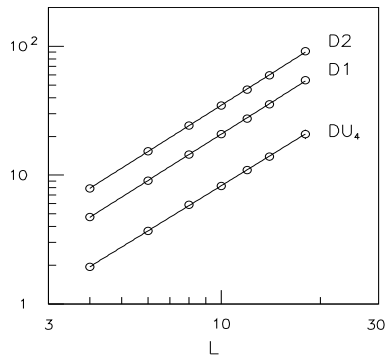}
}
\end{center}
\caption{Determination of the critical exponent $\nu$ from the size dependence of the maxima of thermodynamic response functions. From top to bottom: $\partial \ln \langle m^+{}^2\rangle/\partial T$ (D2), $\partial \ln M/\partial T$ (D1), $\partial U_4/\partial T$ (DU$_4$). The error bars are smaller than the size of symbols.}
\label{nu}
\end{figure}

The critical exponent $\nu$ can be determined independently of any other critical quantity, and therefore more accurately. It can be shown that the slope of the cumulant $U_4$ at the critical temperature $T_c$, or at any other point in the critical region, apart from higher order corrections, scales with system size as $L^{1/\nu}$. \cite{binder} 
The same scaling behavior is exhibited by the logarithmic derivative of any power of the staggered magnetization. We calculated the derivative of $U_4$, of $\ln M$ and of $\ln \langle m^+{}^2 \rangle$ with respect to $T$, using Eq. (\ref {deriv_log}). Figure \ref{nu} displays a log-log plot of the maximum of these derivatives vs. $L$. 
As expected, they show a linear behavior and they are to a very good approximation parallel. No indication of correction to scaling are evident. The slopes found after a linear fit to the data are 1.579 $\pm$ 0.056, 1.619 $\pm$ 0.031, 1.617 $\pm$ 0.028, respectively. The weighted average of these values gives $\nu= 0.620 \pm 0.008$, which, within the error bars, is in reasonable agreement with the Ising value $0.6295 \pm 0.0009$ found by MC simulations, \cite{ising_mc,luijten} but is quite different from the theoretically expected Fisher renormalized $\nu ' = \nu/(1-\alpha) \simeq$ 0.702.  

Various thermodynamic quantities exhibit an extreme at a certain temperature $T_c(L)$, which depends strongly on that quantity and on the system size. In the asymptotic regime of large systems the following scaling law holds
\begin{equation}
T_c(L) \sim T_c + A_x L^{-1/\nu},
\label{tc_scal}
\end{equation}
where the subscript $x$ indicates that the prefactor $A$ depends on the quantity considered. 
Once $\nu$ is determined, Eq. (\ref{tc_scal}) enables us to extrapolate $T_c (L)$ to the thermodynamic limit $L\rightarrow \infty$. 
\begin{figure}[h]
\begin{center} 
\mbox{
\epsfxsize=6cm 
\epsfysize=5cm 
\epsffile{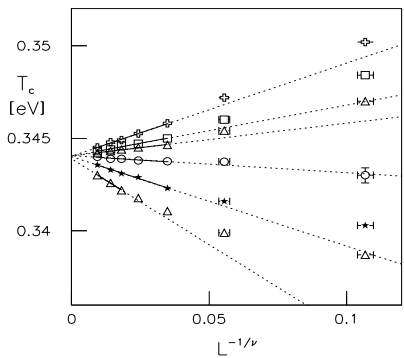}
}
\end{center}
\caption{Extrapolations of the finite system \lq\lq critical temperature\rq\rq to the thermodynamic limit for $\mu_B=1.42$ eV. From top to bottom: $\partial U_4/\partial T$, $\partial \ln \langle m^+{}^2\rangle/\partial T$, $\partial \ln M/\partial T$, $\chi$, $\partial M/\partial T$, and $C_v$. Full lines are fits of Eqs. (\ref{tc_scal}) to the corresponding data. Dotted lines are just extensions of the full ones. The error bars if not visible are smaller than the size of symbols.}
\label{tc}
\end{figure}
In Fig. \ref{tc} we have plotted $T_c(L)$ of the various quantities examined against $L^{-1/\nu}$. The data follow very close the linear behavior expected, except the two smaller lattice sizes $L=4, 6$ for which corrections to scaling are required. We have therefore excluded those data from the extrapolations. The full lines in Fig. \ref{tc} are linear fits that account for both the errors on $x$ and $y$ coordinates. The intercepts on the $T$-axis ($L=\infty$) are very close to each other, however, to account for the slight deviations, we have weight averaged the extrapolated values. The final estimation is $T_c = 0.34404 \pm 0.00006$ eV. 

The critical exponents $\gamma/\nu$ and $\beta/\nu$ can be directly determined from the finite scaling of $\chi^+$ and $M$, respectively at $T_c$. In the asymptotic regime these quantities scale as
\begin{eqnarray}
\chi^+ & \sim & L^{\gamma/\nu}
\label{chi_gamma}
\\
M  & \sim & L^{-\beta/\nu}. 
\label{m_beta}
\end{eqnarray}
\begin{figure}[h]
\begin{center} 
\mbox{
\epsfxsize=6cm 
\epsfysize=5cm 
\epsffile{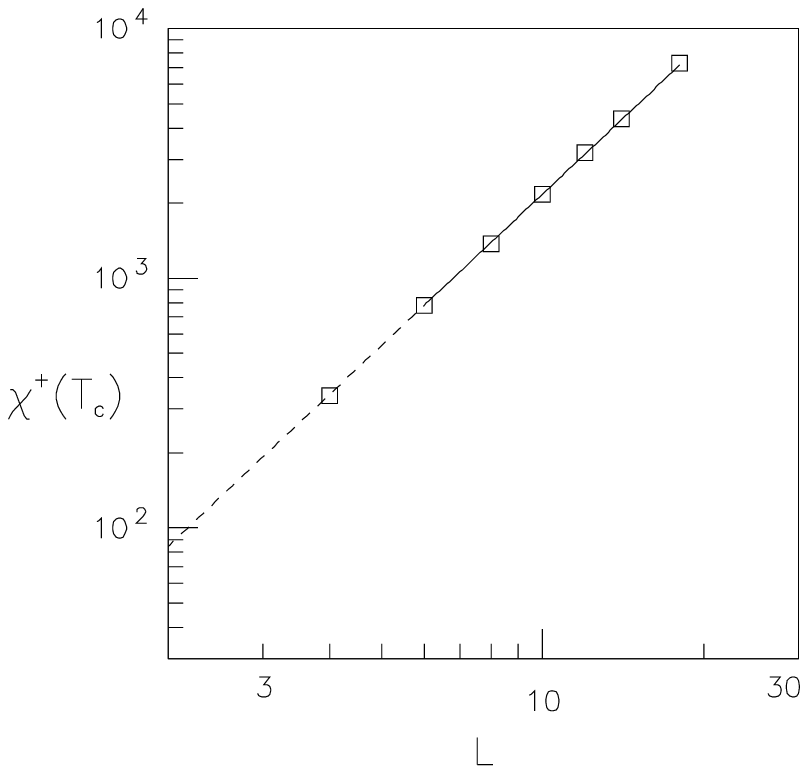}
}
\mbox{
\epsfxsize=6cm 
\epsfysize=5cm 
\epsffile{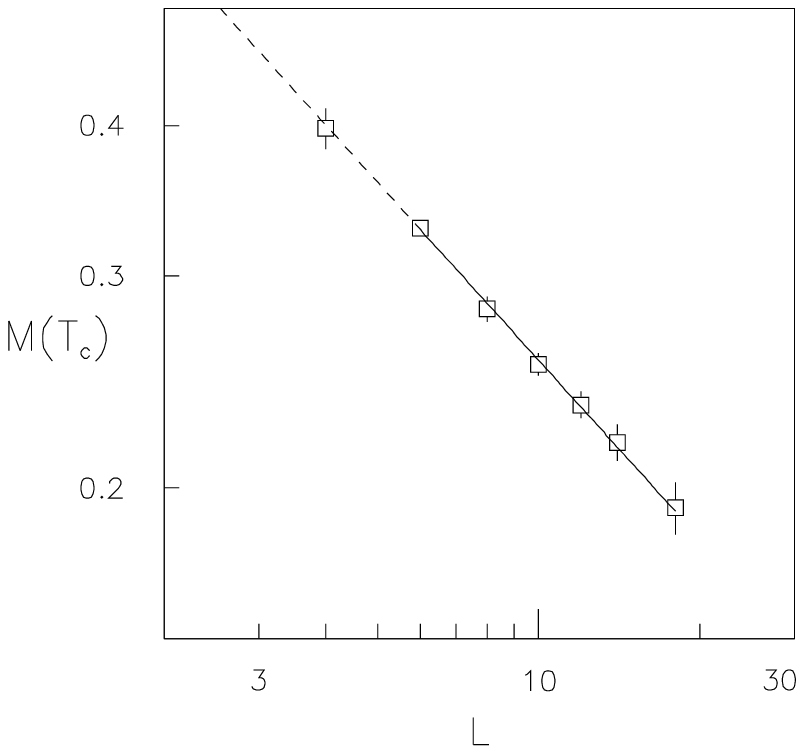}
}
\end{center}
\caption{Determination of the critical exponents by finite scaling relations. (a) Linear fit of $\chi^+(T_c)$ (Eq. (\ref{chi_gamma})), which provides $\gamma/\nu$. (b) Linear fit of $M(T_c)$ (Eq. (\ref{m_beta})), which provides $\beta/\nu$. Full lines are fits, dotted lines are just extensions of the full ones. The error bars if not visible are smaller than the size of symbols.}
\label{gamma_beta}
\end{figure}
Figure \ref{gamma_beta} shows log-log plots of the data. If, again, we exclude the lowest lattice size, the data are found to follow a straight line very well. From a linear fit to the susceptibility we get $\gamma/\nu = 2.017 \pm 0.041$. Using the value previously found for $\nu$, we get $\gamma = 1.25 \pm 0.04$, which, within the errors, is in agreement with the Ising value $\gamma_{\rm {Ising}}= 1.2390 \pm 0.0071$ determined by $\epsilon$-expansion, \cite{eps_ising} and 1.237 $\pm$ 0.002 by MC simulations. \cite{ising_mc,luijten} Analogously, we find $\beta/\nu= 0.491 \pm 0.057$, or $\beta$ = 0.305 $\pm$ 0.039, which, within the errors, agrees with the Ising value $\beta_{\rm{Ising}} = 0.3270 \pm 0.0015$ \cite{eps_ising} as well. We have also tried to determine the critical exponent of the specific heat $\alpha$. This is however, more difficult to measure because of the presence of an additional fitting parameter. The specific heat is, indeed, expected to scale at $T_c$ as
\begin{equation}
C_v \sim B_1 + B_2 L^{\alpha/\nu}.
\end{equation}
The fit of $B_1, B_2$, and $\alpha/\nu$, not shown here, gives $\alpha/\nu = 0.28 \pm 0.09$, or $\alpha$ = 0.17 $\pm$ 0.06, which is in reasonable agreement with the Ising value $\alpha_{\rm{Ising}}$ = 0.110 $\pm$ 0.002. \cite{luijten,alpha_ising} 

The fourth order cumulant $U_4$ is an important quantity to determine the kind of a phase transition and also to provide an independent determination of the critical temperature. The curves $U_{4,L}(T)$ plotted for different $L$ vs $T$ for large $L$ all cross at $T_c$. \cite{binder} Moreover, the value $U_4(T_c)$ strongly depends on the kind of transition. It has been found that $U_4 \simeq 0.27052$ for the mean-field universality class \cite{u4_mf}, $U_4 \simeq 0.47$ for the three-dimensional Ising model, \cite{ising_mc} and $U_4 \simeq 0.5$ for a first order transition. \cite{u4_fo} If the asymptotic regime has not yet entered, however, curves with different $L$ will cross at different points. Nevertheless, it is still possible to extrapolate the crossing point to $L\rightarrow \infty$. The procedure is described in Ref. \onlinecite{ising_mc}. 
\begin{figure}[h]
\begin{center} 
\mbox{
\epsfxsize=6cm 
\epsfysize=5cm 
\epsffile{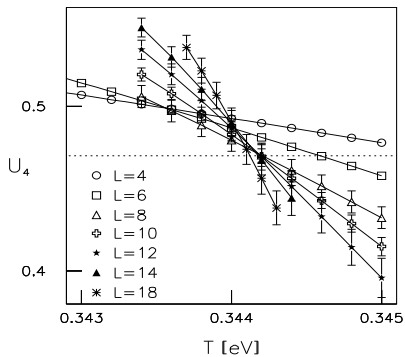}
}
\end{center}
\caption{Plot of the fourth-order cumulant $U_4$ vs $L$. The horizontal dotted line indicates the crossing point value of the rigid Ising model. The full lines are just a guide for the eyes. The error bars if not visible are smaller than the size of symbols.}
\label{u4}
\end{figure}
Figure \ref{u4} displays the behavior of $U_4$ for our system. For $L \gtrsim 8$ the different curves cross with very good approximation at the Ising value. The value obtained for the critical temperature using this method is consistent with the value previously determined. 

\begin{figure}[h]
\begin{center} 
\mbox{
\epsfxsize=6cm 
\epsfysize=5cm 
\epsffile{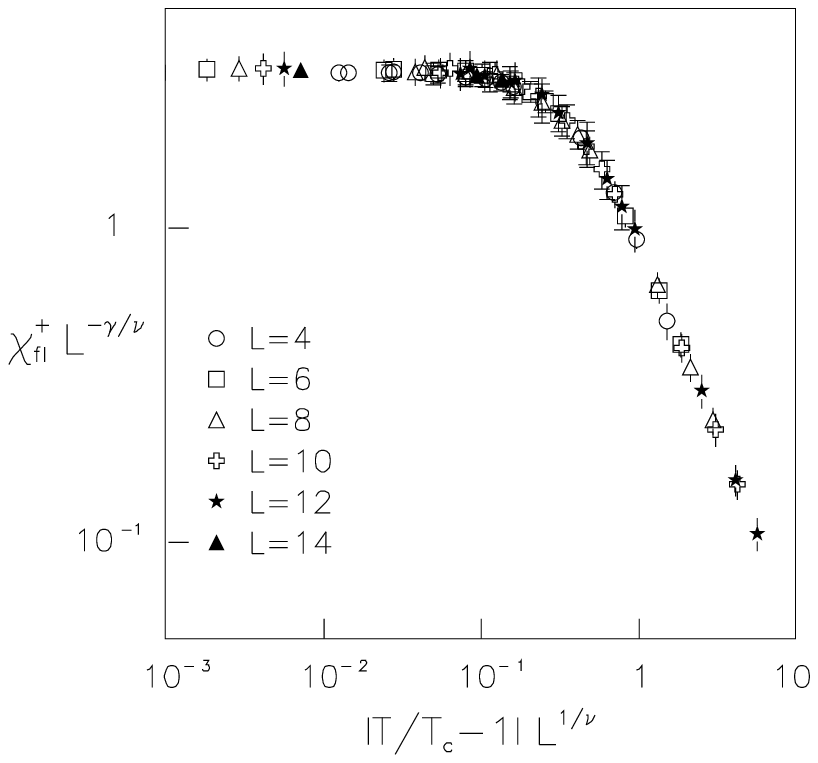}
}
\mbox{
\epsfxsize=6cm 
\epsfysize=5cm 
\epsffile{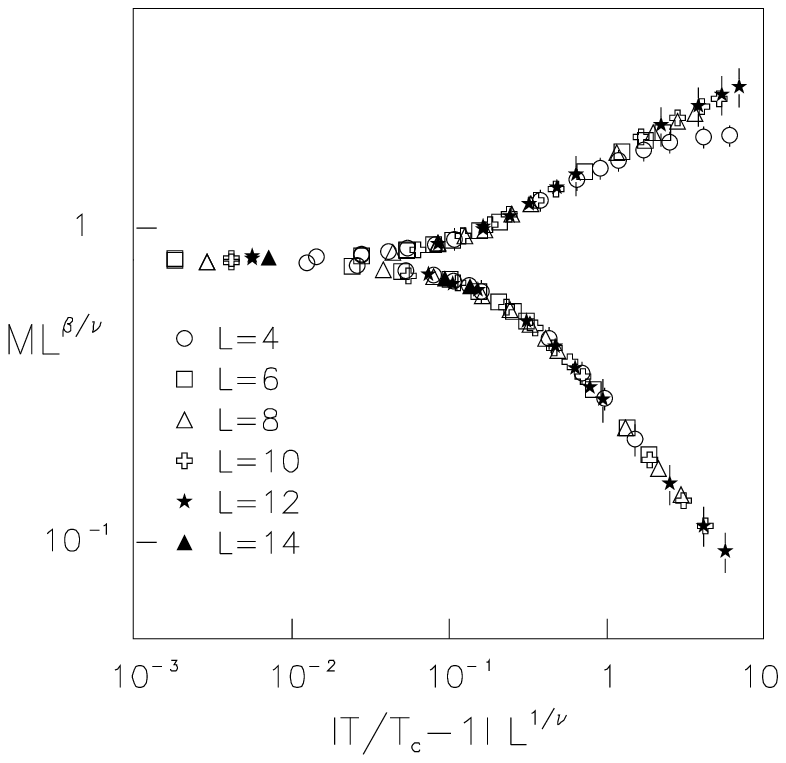}
}
\end{center}
\caption{Data collapsing of the rescaled susceptibility $\chi^+_{fl}$ (a), and of the order parameter $M$ (b) vs rescaled temperature. The error bars if not visible are smaller than the size of symbols.}
\label{fscal}
\end{figure}

For a $d$-dimensional system for which the hyperscaling relation $d\nu = 2 -\alpha$ is valid, the following finite-size laws hold in the vicinity of the critical point 
\begin{eqnarray}
\chi^+_{fl}(L,T) & = & L^{\gamma/\nu} f(t L^{1/\nu}), \\
M  & = & L^{-\beta/\nu}g(t L^{1/\nu}),
\end{eqnarray}
where $t=1-T/T_c$. In a scaling plot of $\chi^+_{fl} 
L^{-\gamma/\nu}$ (resp. $ M L^{\beta/\nu}$) vs $|1-T/T_c| L^{1/\nu}$ one should, therefore, observe a collapsing of the data. This is exactly what we found, as Fig. \ref{fscal} demonstrates, and is a further evidence of the consistency of the critical exponents and 
$T_c$ previously determined.

\section{conclusions}
We have performed Monte Carlo simulations of the compressible Ising model with antiferromagnetic interactions under constant volume conditions, in the semi-grand-canonical ensemble. Elastic forces are included by the Stillinger-Weber potential. The behavior of all critical quantities analyzed strongly indicated the presence of a closed second order line with the critical exponents of the (rigid) Ising model. This is in contrast with theories as they predict the occurrence of Fisher renormalized exponents. Disagreement was also found in the simulations of exactly the same model but with ferromagnetic interactions \cite{francesca}. The reasons of these disagreements are not clear and should be further investigated. Needless to say, however, that our conclusions should be viewed within the context of any numerical work, keeping in mind the finiteness of the systems used in the simulations. It is, therefore, clear that, on the basis of these data, the occurrence of a (slow) crossover toward Fisher renormalized exponents cannot be completely ruled out. A deeper investigation of this issue would require simulations on much larger system sizes, which is, basically, unfeasible with the present computer power. It would be, however, very interesting to investigate the critical behavior under stronger coupling conditions.
\section{Acknowledgments}
We thank B. D\"unweg, F. Tavazza, and X. Zhu for helpful comments and discussions. The support of NSF Grant No. DMR-0341874 is gratefully acknowledged.


\begin{thebibliography}{99}
\bibitem{ipotesi_burk}The following assumptions are made: (i) Only nearest neighbor AFM interactions. (ii) Isotropic elasticity, i.e. only two elastic constants are present.
\bibitem{nota_afcv}This conclusion is valid if $\alpha >0$, and at strong coupling. If $\alpha < 0$ Ising exponents are predicted.
\bibitem{nota_afcp}This conclusion is valid if $\alpha >0$. If $\alpha < 0$ Ising exponents are predicted.
\bibitem{nota_t}In the present work, unless otherwise specified, the temperature is quoted as $k_B T$. This permits easy comparison with energy scales used in electronic structure.
\end{thebibliography}
\end{document}